\documentclass[prb,twocolumn,superscriptaddress,amsmath,amssymb]{revtex4}
\pdfoutput=1
\usepackage{natbib}
\usepackage{graphicx}         
\usepackage{dcolumn}            
\usepackage{bm}               
\usepackage{booktabs}
\usepackage{tabularx}

\begin{document}

\preprint{}

\title{Supercurrent in Nb/InAs-Nanowire/Nb Josephson junctions}

\author{H. Y. G\"unel}

\affiliation{Peter Gr\"unberg Institute (PGI-9) and JARA-Fundamentals of Future Information  Technology, Forschungszentrum J\"ulich GmbH, 52425 J\"ulich, Germany}

\author{I. E. Batov}

\affiliation{Institute of Solid State Physics, Russian Academy of Sciences, Chernogolovka, 142432 Moscow district, Russia}

\author{H. Hardtdegen}

\author{K. Sladek}

\author{A. Winden}

\author{K. Weis}

\affiliation{Peter Gr\"unberg Institute (PGI-9) and JARA-Fundamentals of Future Information  Technology, Forschungszentrum J\"ulich GmbH, 52425 J\"ulich, Germany}

\author{G. Panaitov}

\affiliation{Peter Gr\"unberg Institute (PGI-8) and JARA-Fundamentals of Future Information  Technology, Forschungszentrum J\"ulich GmbH, 52425 J\"ulich, Germany}

\author{D. Gr\"utzmacher }

\affiliation{Peter Gr\"unberg Institute (PGI-9) and JARA-Fundamentals of Future Information  Technology, Forschungszentrum J\"ulich GmbH, 52425 J\"ulich, Germany}

\author{Th. Sch\"apers}
\affiliation{Peter Gr\"unberg Institute (PGI-9) and JARA-Fundamentals
of Future Information  Technology, Forschungszentrum J\"ulich GmbH,
52425 J\"ulich, Germany} \affiliation{II. Physikalisches Institut, RWTH
Aachen University, Aachen, Germany}

\date{\today}

\hyphenation{}

\begin{abstract}
We report on the fabrication and measurements of planar mesoscopic
Josephson junctions formed by InAs nanowires coupled to superconducting
Nb terminals. The use of Si-doped InAs-nanowires with different bulk
carrier concentrations allowed to tune the properties of the junctions.
We have studied the junction characteristics as a function of
temperature, gate voltage, and magnetic field. In junctions with high
doping concentrations in the nanowire Josephson supercurrent values up
to 100\,nA are found. Owing to the use of Nb as superconductor the
Josephson coupling persists at temperatures up to 4\,K. In all
junctions the critical current monotonously decreased with the magnetic
field, which  can be explained by a recently developed theoretical
model for the proximity effect in ultra-small Josephson junctions. For
the low-doped Josephson junctions a control of the critical current by
varying the gate voltage has been demonstrated. We have studied
conductance fluctuations in nanowires coupled to superconducting and
normal metal terminals. The conductance fluctuation amplitude is found
to be about 6 times larger in superconducting contacted nanowires. The
enhancement of the conductance fluctuations is attributed to
phase-coherent Andreev reflection as well as to the large number of
phase-coherent channels due to the large superconducting gap of the Nb
electrodes.
\end{abstract}

\maketitle

\section{Introduction}

Coherent transport in mesoscopic semiconductor-based Josephson
junctions has been attracting a lot of interest from fundamental and
applied physics point of view. Modern nanofabrication techniques
provide a possibility to fabricate planar multi-terminal Josephson
structures relevant for the realization of different types of hybrid
superconductor/semiconductor nanoscale
devices.\cite{DeFranceschi10,Schaepers01} Josephson field effect
transistors,\cite{Akazaki96} superconducting quantum point contacts
\cite{Hideaki95} and injection current controlled Josephson junctions
\cite{Schaepers03a} were realized with high-mobility 2-dimensional
electron gases in semiconductor heterostructures. Recently, the
Josephson effect was also observed in nanoscale devices formed by
semiconductor nanowires coupled to superconducting terminals.
\cite{Doh05,Xiang06,vanDam06,Frielinghaus10,Roddaro11,DeFranceschi10,Nishio11}

Up to now, most of experimental studies of semiconductor-based
Josephson junctions at nano\-scale have been focused on structures with
an InAs-nanowire as a semiconductor weak link. Experiments show that a
charge accumulation layer is formed at the surface of the InAs,
\cite{Smit89} which  provides a sufficiently low resistive contact to
superconducting electrodes. It was found that highly transparent
contacts are formed at the interfaces between the InAs nanowire and Al
superconducting electrodes. At temperatures below 1~K the high
transparency of the contacts gives rise to proximity-induced
superconductivity. A large number of experiments have been carried out
on Al/InAs-nanowire junctions, demonstrating tunable Josephson
supercurrents\cite{Doh05,Xiang06}, supercurrent
reversal,\cite{vanDam06} Kondo enhanced Andreev
tunneling,\cite{Sand-Jespersen07} and suppression of supercurrent by
hot-electron injection.\cite{Roddaro11} Furthermore, gate-controlled
superconducting quantum interference devices,
\cite{vanDam06,DeFranceschi10} and tunable Cooper pair
splitters\cite{Hofstetter09} have been realized.

The use of Al as a superconductor in nanowire-based Josephson junctions
limits the operation temperature of the nanofabricated devices below $T
\approx1.2$\,K. To extend the operation of the Josephson devices at
higher temperatures, it has been suggested to use superconductors which
have higher temperatures of superconducting transition. Recently,
Spatnis \emph{et al.} \cite{Spathis11} have realized proximity dc
SQUIDs based on InAs nanowires and vanadium superconducting electrodes
($T_c \approx 4.6$\,K). It is shown that V/InAs-nanowire/V Josephson
junctions can operate at temperatures up to 2.5\,K. Furthermore, for
InN-nanowire-based junctions with Nb electrodes ($T_c\approx 9.3$\,K) a
Josephson supercurrent was observed up to temperatures of
$3.5$\,K.\cite{Frielinghaus10} It is also pointed out that, owing to
the strong spin-orbit coupling in InAs nanowires,\cite{Fasth07} the
mesoscopic devices fabricated could be used for the experimental
demonstration of Majorana fermions \cite{Lutchyn10}.

In some aspects small-size planar superconductor/normal
conductor/superconductor Josephson junctions differ significantly from
their large-size counterparts, i.e. the magnetic field dependence of
the critical current $I_c$ is expected to decrease mono\-tonously with
increasing magnetic field $B$ in contrast to the Fraunhofer-type $I_c$
vs. $B$ dependence for larger junctions.\cite{Cuevas07,Bergeret08}
Indeed, a monotonous decrease of $I_c$ was recently observed in
Nb/Au/Nb and Al/Au/Al junctions as well as in  Nb/InN-nanowire/Nb
junctions.\cite{Angers08,Frielinghaus10} Furthermore, in junctions with
a semiconductor nanowire in between two superconducting electrodes
mesoscopic effects, i.e. conductance fluctuations and fluctuations of
the critical current have been observed when the gate voltage is
varied.~\cite{Doh05,Doh08,Jespersen09} The gate voltage dependent
conductance fluctuations measured at finite bias voltage were shown to
follow almost precisely the fluctuations of the supercurrent
\cite{Takayanagi95,Doh05}.  It was also found that the conductance
fluctuation amplitude measured at bias voltages lower than 2$\Delta$/e
significantly exceeds the amplitude of the normal-state universal
conductance fluctuations.\cite{Doh08,Jespersen09}

In our study we investigated the Josephson effect in
Nb/InAs-nanowire/Nb hybrid structures. By using InAs-nanowires with two
different bulk carrier concentrations we were able to vary the
properties of the nanoscale weak link Josephson junction to a large
extent, i.e. in the junctions with the highly doped InAs nanowires a
relatively large Josephson supercurrent is observed, whereas for
junctions with low doped nanowires $I_c$ is reduced but here a gate
control of the superconducting switching currents is possible. We
devoted special attention to effects which originate from the small
size of the junctions, in particular the magnetic field dependence of
the critical current as well as the gate voltage dependent conductance
and supercurrent fluctations. The field dependence of the critical
current and the fluctuation phenomena are compared to recent
theoretical models.

\section{Experimental}

\begin{figure}[h!]
\begin{center}
\includegraphics[width=1.0\columnwidth]{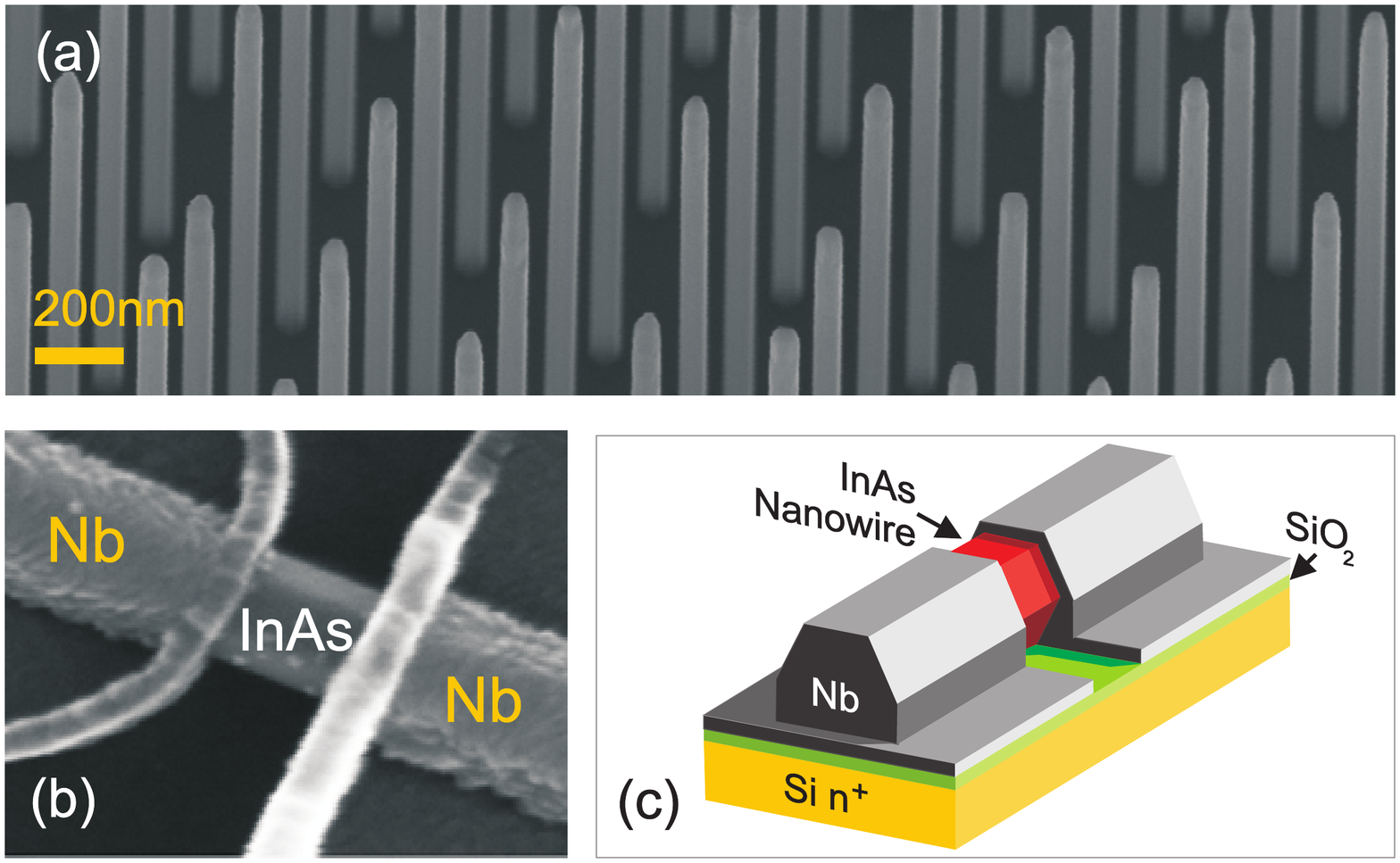}
\caption{(a) Scanning electron micrograph of the as grown nanowires
with doping factor 500. (b) scanning
electron micrograph of sample L2. (c) Schematic illustration of the junction layout.}\label{fig:1}
\end{center}
\end{figure}

The $n$-type doped InAs nanowires were grown by selective area metal
organic vapor phase epitaxy (MOVPE) without using catalyst material. In
order to tune the Si doping level, the ratio of disilane
(Si$_{2}$H$_{6}$) partial pressure and group III precursor has been
adjusted as
$p(\mathrm{Si}_{2}\mathrm{H}_{6})/p(\mathrm{TMIn})=7.5\times 10^{-5}$,
which is defined as doping factor 1. In this study we have used
nanowires with two different doping concentrations.\cite{Wirths11} The
resistivity for the lower (doping factor 100, carrier concentration
$n_{3d}\approx 1 \times 10^{18}$cm$^{-3}$) and higher doped (doping
factor 500, $n_{3d}\approx 7 \times 10^{18}$cm$^{-3}$) nanowires was
0.019 and 0.0018~$\Omega$cm, respectively. In Fig.~\ref{fig:1}(a) a
scanning electron micrograph of the as-grown nanowires is shown.
Detailed information concerning the growth and characterization of the
doped nanowires can be found elsewhere.\cite{Wirths11} The samples are
labeled as "L1, L2" and "H1, H2" for junctions based on a nanowire with
lower (doping factor 100) or higher (doping factor 500) doping
concentration, respectively.

In order to contact the nanowires with Nb electrodes they were
transferred from the growth substrate to an $n^+$-Si/SiO$_{2}$
substrate with predefined electron beam markers. The Nb electrodes were
defined by electron beam lithography and lift-off. Before Nb deposition
the samples were exposed to an oxygen plasma to remove electron beam
resist residues on the contact area. Furthermore, in order to obtain a
high contact transparency, Ar$^{+}$ ion milling was employed to remove
the native oxide on the nanowire surface. The 100~nm thick Nb layer was
deposited by sputtering. The highly doped $n$-type substrate was used
for back-gating of the nanowires. A scanning electron micrograph of a
typical junction (sample L2) is shown in Fig.~\ref{fig:1}(b), while  a
schematics of the junction lay-out can be found in Fig.~\ref{fig:1}(c).
The contact separations $L$ of the four different samples are given in
\ref{tab:1}. The transport measurements were performed at temperatures
down to 0.3\,K in a He-3 cryostat equipped with a superconducting
solenoid with a magnetic field up to 7\,T. The DC and differential
current-voltage characteristics were measured using a four-terminal
current-driven measurement scheme. Current and voltage leads were
filtered by RC filters thermally anchored at $\sim$2 K. The
differential resistance $dV/dI$  was measured with a lock-in amplifier
by superimposing a small AC signal of 5\,nA to the junction bias
current.

\section{Results and Discussion}

\begin{figure}[h!]
\begin{center}
\includegraphics[width=0.8\columnwidth]{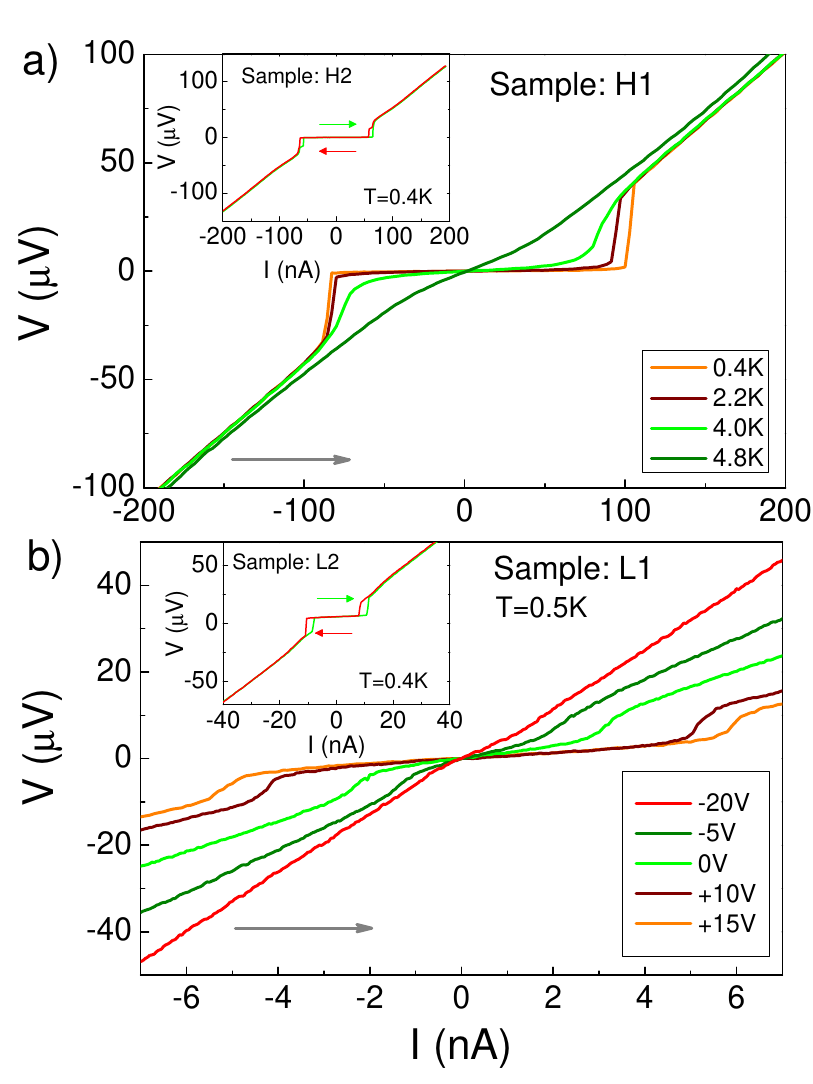}
\caption{(a) $IV$-characteristics of a highly doped
device (sample H1) at various temperatures. The inset shows the
$IV$-characteristics of the second highly doped device
(sample H2) at 0.4~K.  (b) Gate voltage dependent $IV$-characteristics of the junction with the low doped nanowire
(sample L1) at 0.5~K, the upper inset shows the $IV$-characteristics of the second device with a low-doped nanowire
(sample L2) at zero gate voltage. The arrows in (a) and (b) indicate the sweep direction.}\label{fig:2}
\end{center}
\end{figure}

In Fig.~\ref{fig:2}(a) the current-voltage ($IV$) characteristics at
various temperatures between 0.4 and 4.8\,K are shown for a sample with
a highly doped nanowire (sample H1). At temperatures $T\leq 4$\,K and
small bias, a clear Josephson supercurrent is observed in the junction.
As the bias current exceeds a certain value $I_{sw}$, the Josephson
junction switches from the superconducting to the normal state. The
switching current $I_{sw}$ measured at 0.4\,K for the sample H1 is
about 100\,nA. With increasing the temperature $I_{sw}$ is reduced. At
$T>4.5$\,K the supercurrent is suppressed completely. Our measurements
show that at temperatures below 2\,K the $IV$-characteristics are
hysteretic. With increasing the temperature the hysteresis is gradually
suppressed. The retrapping current $I_r$ , defined by the switching
from the normal state back into the superconducting state, has a value
of 84\,nA at 0.4\,K and remains almost constant when the temperature is
increased to 2\,K. As can be inferred from the $IV$-characteristics
shown in Fig.~\ref{fig:2}(a) (inset) the switching current $I_{sw}$ of
the second sample with the highly doped nanowire (sample H2) has a
somewhat smaller value of 70\,nA compared to the first sample.

It should be noted that the switching current $I_{sw}$ is strongly
affected by the external on-chip $RC$ circuit integrated with the
Josephson junction\cite{Jarillo-Herrero06,Pallecchi08,Jespersen09} and
depends on the quality factor $Q$ of the devices. Following the
approach given in Refs.\cite{Jarillo-Herrero06,Jespersen09} we have
estimated the quality factor $Q$ of the samples H1, H2 taking into
account the electromagnetic environment. Estimations show that $Q < 1$,
thus, the junctions H1, H2 are in the overdamped limit and the
measurable supercurrent $I_{sw}$ approaches the thermodynamic critical
current $I_c$. The observed hysteresis in the $IV$ characteristics in
our planar mesoscopic SNS Josephson junctions can be explained by the
increase of the electron temperature in the weak link once the junction
switches to the resistive state.\cite{Courtois08}

Due to the lower electron concentration, gate control was achieved for
the samples with the lower doped nanowires. For sample L1 this is shown
in Fig.~\ref{fig:2}(b), where the $IV$-characteristics at 0.5~K are
plotted for back-gate voltages between $-20$~V and $+15$~V. At zero
gate voltage the measured $I_c$ is about 2.8~nA, which is considerably
smaller than the $I_c$ of the junctions with the highly doped
nanowires. We attribute the lower $I_c$ to the larger resistivity of
the lower doped nanowires. As can be seen in Fig.~\ref{fig:1}(b), by
applying a back-gate voltage of $+15$~V the switching current can be
increased by about a factor of 2. The larger value of the switching
current is due to the fact that by applying a positive gate voltage the
electron concentration in the nanowires is increased so that the
resistivity is decreased. In contrast, by applying a negative gate
voltage the electron concentration in the nanowire is reduced. As a
consequence, $I_c$ is reduced [cf. Fig.~\ref{fig:2}(b)], and at a gate
voltage of $-20$~V the supercurrent is suppressed completely. As can be
seen in Fig.~\ref{fig:2}(b), a small voltage drop appears for currents
below $I_c$ which can be attribute thermal smearing. A Josephson
supercurrent was also observed for the second sample with a low doped
nanowire (sample L2) [cf. Fig.~\ref{fig:2}(b) (inset)]. Here, $I_c$ is
found to be 12~nA at 0.4~K and thus larger than the value for sample
L1. As discussed in detail below, for sample L2 a gate control of $I_c$
was obtained as well.

\begin{table}\centering
\caption{Sample parameters: The contact separation, $L$, and the
diameter of nanowires, $d$, have been determined from scanning electron
micrograph; $I_{c}$ is the critical current and $R_{N}$ is the normal
state resistance of the samples. \label{tab:1}}
\begin{tabular}{l c c c c}\hline\hline
\multicolumn{1}{l}

Device & $L$(nm)  & $d$(nm) & $R_{N}$(k$\Omega$) & $I_{c}$(nA)\\
\hline
Sample H1  &  140  &  110  &  0.75  &  100\\
Sample H2  &  160  & 100  &  0.90  &  70\\
Sample L1  &  70  & 80  &  3.8  &  2.8\\
Sample L2  &  85  & 75  &  2.5  &  12\\
\hline
\end{tabular}
\end{table}

\begin{figure}[h!]
\begin{center}
\includegraphics[width=1.0\columnwidth]{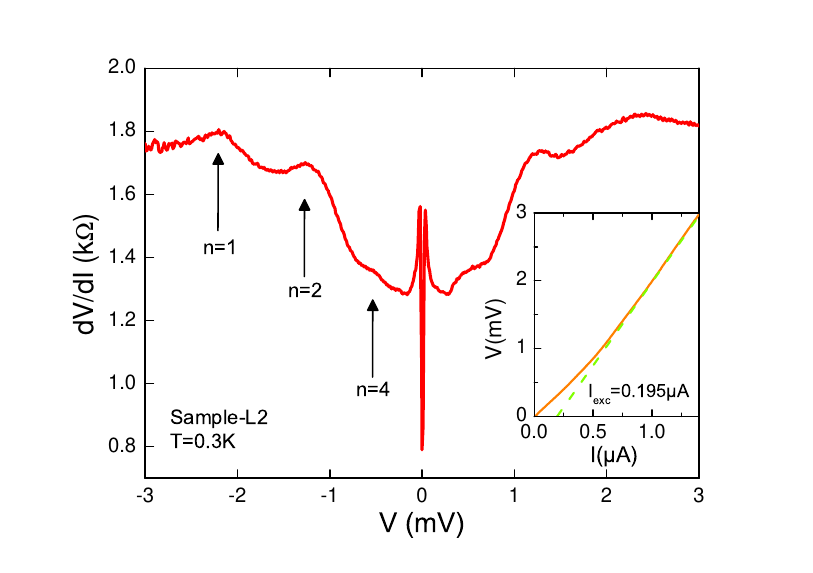}
\caption{Differential resistance $dV/dI$ versus bias voltage for sample
L2 at 0.3\,K. The subgap features are
indicated by arrows. The inset shows a
high bias range $IV$-characteristic of
the device (solid line) and the linear fit of
the $IV$-curve at $V>2 \Delta$ (dashed line)
to show the excess current value
$I_{exc}$.} \label{fig:3}
\end{center}
\end{figure}

In Fig.~\ref{fig:3}, the differential resistance $dV/dI$ vs bias
voltage of the sample L2 is shown. The measurement temperature was
0.3~K. The peaks in $dV/dI$ at finite bias voltages can be associated
with the subharmonic energy gap structure caused by multiple Andreev
reflections, with peak positions given by $V_{n}=2\Delta/en$ $(n= 1,
2,\ldots)$.\cite{Octavio1983,Flensberg88,Cuevas06} From the fit of the
peak positions we have determined the superconducting energy gap
$\Delta$ in the Nb leads $\Delta=1.2$\,meV and found that the observed
peaks correspond to $n=1, 2,$ and 3. The contact transparency in the
device L2 can be estimated using the $IV$-characteristic for $V>2
\Delta/e$.\cite{Xiang06,Nishio11} We have found that the fit to the
$\emph IV$-curve at the normal state in the range $V>2.5$\,mV
extrapolates to a finite excess current $I_{exc}=195$\,nA as shown in
the inset of Fig.~\ref{fig:2}. Using the superconducting energy gap
$\Delta=1.2$\,meV and the normal state resistance of the junction
$R_{n}=2.47$\,k$\Omega$, we obtain $eI_{exc}R_{n}/\Delta$=0.4. We
analyzed the data within the framework of the standard
Blonder-Tinkham-Klapwijk (BTK) theory.\cite{Blonder82} The obtained
value $eI_{exc}R_{n}/\Delta=0.4$ is converted to the BTK barrier
strength parameter $Z \approx 0.85$,\cite{Flensberg88} which
corresponds to a contact transparency $T_{n} \approx 0.6$. Subgap
features as found here have been observed before in other
superconductor-semiconductor
junctions.\cite{Frielinghaus10,Batov04,Chrestin97} From the
electron-phonon coupling strength 2$\Delta
_{0}$/k$_{B}$T$_{c}$=3.9\cite{Carbotte90,Frielinghaus10} and assuming
$\Delta _{0}$=1.2~meV for the superconducting gap at $T=0$, we obtain a
critical temperature $T_c=7.2$\,K for the Nb electrodes, which fits
well to the measured value of $T_c$.

\begin{figure}[h!]
\begin{center}
\includegraphics[width=0.8\columnwidth]{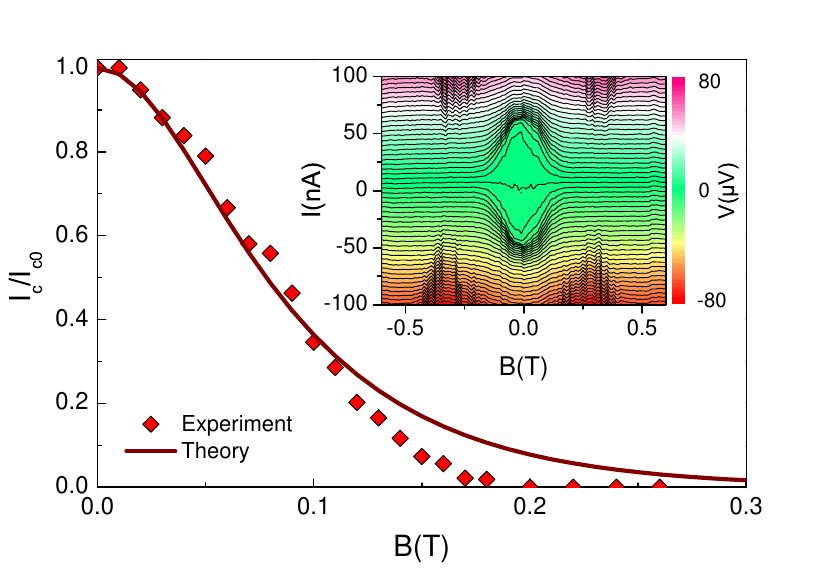}
\caption{Normalized critical current
$I_c/I_{c0}$ versus magnetic field $B$
($\diamond$) for sample H2. The solid line
represents the calculation  according to
the theoretical model. \cite{Cuevas07,Bergeret08}
The inset shows the color-scaled voltage drop as a function of magnetic field
and bias current. The contour lines correspond to constant voltage values separated by $2.5\,\mu$V.}
\label{fig:4}
\end{center}
\end{figure}

Owing to the large critical field of Nb a Josephson supercurrent is
maintained up to relatively large magnetic fields. This is illustrated
in Fig.~\ref{fig:4} (inset), where the color-scaled voltage drop
measured at sample H2 is given as a function of magnetic field and bias
current. The measurement temperature was 0.4~K. The magnetic field was
applied perpendicular to the substrate. The resulting values of $I_c$
normalized to the zero-field critical current $I_{c0}$ are plotted as a
function of $B$ in Fig.~\ref{fig:4}. A monotonous decrease of the
measured critical current $I_c$ with magnetic field $B$ is found. A
complete suppression of $I_c$ occurred at about 0.2\,T. Earlier, a
similar behavior of the critical current with magnetic field was
observed in planar Nb/Au/Nb and Al/Au/Al Josephson junctions and in InN
nanowire-based Josephson junctions.\cite{Frielinghaus10} The monotonous
decrease of $I_c$ with increasing $B$ can be explained within the
framework of a recently developed theoretical model for the proximity
effect in diffusive narrow-width Josephson
junctions.\cite{Cuevas07,Bergeret08} There, it is shown that for
junctions with a width comparable to or smaller than the magnetic
length $\xi_{B}=\sqrt{\Phi_{0}/B}$, the magnetic field acts as a
pair-breaking factor that suppresses monotonously the proximity-induced
superconductivity in the wire and the critical current.

For sample H2 the characteristic magnetic field $B_{0}$ defined by the
flux quantum through the cross section of the normal wire,
$B_{0}=\Phi_{0}/Ld$, is as large  0.13~T, resulting in $\xi_B=126$~nm.
The value of $\xi_{B}$ is thus comparable to the junction width, so
that the theoretical model for the limit of narrow-width Josephson
junctions can be applied.\cite{Cuevas07,Bergeret08} We calculated the
expected dependence of $I_c$ on $B$ using the Thouless energy defined
by $E_{Th}=\hbar D/L^{2}$ as a fitting
parameter.\cite{Cuevas07,Bergeret08,Hammer07} Here, $D$ is the
diffusion constant and $L$ is the length of the junction. As can be
seen in Fig.~\ref{fig:4}, a good agreement between experiment and
theory is obtained. The best fit has been achieved for
$E_{Th}^{fit}=0.2$~meV. This value is smaller than $E_{Th}=0.47$~meV,
determined from transport measurements in the normal state. We
interpret the lower value of $E_{Th}^{fit}$ as compared to the Thouless
energy obtained from the transport data by the presence of interface
barriers in the junctions.\cite{Frielinghaus10}

\begin{figure}[h!]
\begin{center}
\includegraphics[width=0.9\columnwidth]{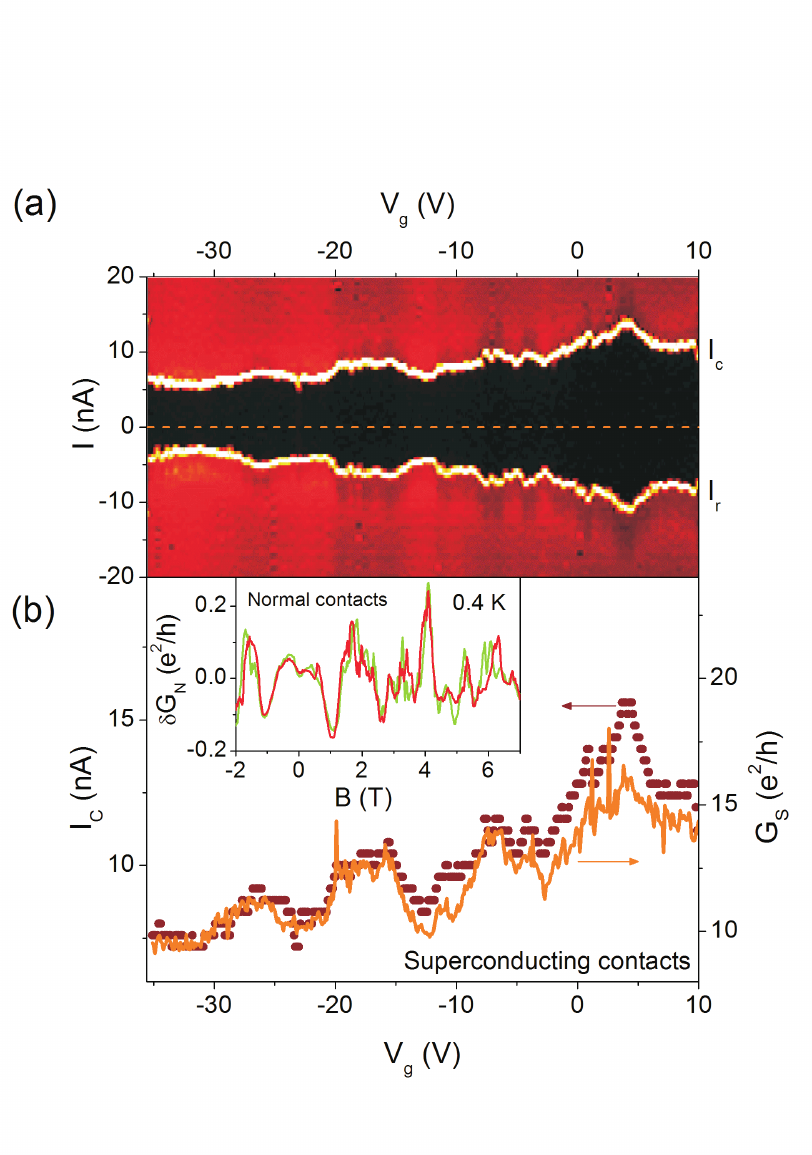}
\caption{(a) Differential resistance
($dV/dI$) plotted as a function of bias current and gate
voltage. (b) Fluctuations of the conductance $G_S$ and the critical
current $I_c$ as a function of gate voltage. The inset graph shows two subsequent measurements of
a normal contacted nanowire.} \label{fig:5}
\end{center}
\end{figure}

In Fig.~\ref{fig:5}(a) the differential resistance $dV/dI$ of sample L2
is plotted in color scale as a function of bias current and back-gate
voltage. The measurements were taken at 0.4~K. It can be seen that on
average the supercurrent range (black region) is reduced if a more
negative gate voltage is applied. This is due to the corresponding
decrease of the electron concentration in the nanowire. Since the
current was biased from negative to positive values, the switching is
nonsymmetric with respect to zero current, with the transition at
negative and positive bias currents corresponding to the return current
$I_r$ and the critical current $I_c$, respectively. A closer look on
Fig.~\ref{fig:5}(a) reveals that $I_c$ fluctuates as a function of
back-gate voltage. The corresponding values of $I_c$ are plotted in
Fig.~\ref{fig:5}(b). The average amplitude of the critical current
fluctuations, i.e.~the root-mean-square (rms) of fluctuations over the
applied gate voltage range, is found to be $\mathrm{rms}(\delta
I_{c})\approx 0.9$~nA. The critical current fluctuations $\delta I_c$
were calculated by subtracting a linear increasing background current.

Mesoscopic fluctuations of the critical current have been theoretically
studied mainly in two different regimes. For the short junction limit
it is found that the fluctuations are universal and that the
fluctuation amplitude depends only on the superconducting gap,
$\Delta$: $\delta I_{c}\sim
e\Delta/\hbar$.\cite{Macedo94,Beenakker91a,Beenakker94} The limit of
long Josephson junctions, where the Thouless energy E$_{Th}$ is much
smaller than the superconducting gap in the leads has been investigated
by Altshuler and Spivak.\cite{Altshuler87} In this model the energy
scale for mesoscopic fluctuations in the critical current, $I_{c}$, is
set by the Thouless energy. Since in our case $E_{Th} \ll \Delta$, we
compared the experimental values of $\mathrm{rms}(\delta I_{c})$ with
the model of Alt'shuler and Spivak which is appropriate for this
particular regime.\cite{Altshuler87} Within this model the amplitude of
supercurrent fluctuations at $T=0$ is given by $\mathrm{rms}(\delta
I_{c})=0.60\; eE_{Th}/\hbar$. By taking a Thouless energy $0.05$\,meV,
as was estimated from the magnetic field dependence of $I_c$ for the
sample L2, we obtain an expected fluctuation amplitude of
$\mathrm{rms}(\delta I_{c}) \approx 7.3$~nA. In the more recent model
of Houzet and Skvortsov,\cite{Houyet08} the proximity effect and the
resulting formation of a minigap in the normal conductor is included in
the analysis of the critical current fluctuations. For the long
junction limit they obtained $\mathrm{rms}(\delta I_{c})=1.49\;
E_{Th}e/\hbar$. According to this model we find an even higher expected
fluctuation amplitude of approximately 18~nA. For both models the
expected values of $\mathrm{rms}(\delta I_{c})$ are considerably larger
than the corresponding experimentally obtained values. The most
probable reason of the lower measured value of $\mathrm{rms}(I_{c})$ is
the presence of a non-ideal superconductor/normal conductor
interface,\cite{Takayanagi95,Doh05} i.e. an interface barrier or
different Fermi velocities in both materials.\cite{Schaepers97} Both
contributions lead to a decrease of $I_c$ and can be expected to result
in an according decrease of $\mathrm{rms}(I_{c})$.

In Fig.~\ref{fig:5}(b) the normalized differential conductance $G_S$ is
plotted in units of $e^2/h$ as a function of back-gate voltage. The
conductance values have been taken at a bias voltage 0.1\,mV, which is
well below $2\Delta/e$, so that multiple Andreev reflections partially
contribute to the total conductance. However, we did not attempt to
drive the junction into a higher bias state above $2\Delta/e$ or
measure the fluctuations at temperatures above $T_c$, such that any
superconducting properties are suppressed. The reason is that under
these conditions the quasi-equilibrium phase coherent transport regime
is left. As can be seen in Fig.~\ref{fig:5}(b), the fluctuation pattern
of $G_S$ follows almost perfectly the pattern of the previously
discussed fluctuations in $I_c$. A similar agreement between the
fluctuation patterns of $I_c$ and $G_S$ has been observed before on
Al/InAs-nanowire and on Nb/2-dimensional electron gas Josephson
junctions.\cite{Doh05,Takayanagi95}

The fluctuations in $G_S$ originate from the phase-coherent transport
through a conductor with small dimensions where only a limited number
of scattering centers are
involved.\cite{Stone85,Altshuler85b,Bloemers11} For our junction we
find a fluctuation amplitude of $\mathrm{rms}(\delta G_S)=1.18
(e^{2}/h)$, with $\delta G_S$ calculated by subtracting the linearly
increasing background conductance. In order to compare
$\mathrm{rms}(\delta G_{S})$ obtained for a sample with superconducting
electrodes with the corresponding value of $\mathrm{rms}(\delta G_{N})$
of a normal conducting reference sample, we have contacted a nanowire
from the same growth run with normal Au/Ti electrodes. Here, the
contact separation was 130~nm. As shown in Fig.~\ref{fig:5}(b) (inset),
at a temperature of 0.4~K reproducible conductance fluctuations are
measured as a function of magnetic field. However, the average
conductance fluctuation amplitude of $\mathrm{rms}(G_{N})=0.2
(e^{2}/h)$ is significantly lower than the value found for the sample
with superconducting electrodes. We attribute the enhanced conductance
fluctuation observed for the sample with superconducting electrodes to
the additional contribution of phase-coherent Andreev reflection. A
similar behavior has been found in other Josephson junctions as
well.\cite{Doh05,Doh08,Trbovic10,Jespersen09,Ojeda-Aristizabal09} In
Al-based Josephson junctions an enhancement of the average fluctuation
amplitude between $1.4$ and $1.6$ has been reported,\cite{Trbovic10}
whereas in our Nb contacted junctions, we have found an enhancement by
a factor of about 6. The larger enhancement compared to the values
reported for Al-based junctions can be qualitatively explained by the
larger superconducting energy gap of Nb compared to Al, resulting in a
larger number of phase-coherent Andreev channels.

\section{Summary and Conclusion}

In summary, we have successfully fabricated and characterized
Nb/InAs-nanowire/Nb Josephson junctions. By taking advantage of Nb as a
superconductor, we could demonstrate that the junctions comprising a
highly doped InAs nanowire show a clear Josephson supercurrent up to
relatively high temperatures of 4~K. For the junctions with the lower
doped nanowire, gate control of the Josephson supercurrent was
achieved. The measurements of $I_{c}$ as a function of magnetic field
show that a Josephson supercurrent can be maintained up to a field of
0.2~T. The observed monotonous decrease of $I_c$ with increasing
magnetic field is explained by the magnetic pair breaking effect in
narrow-width Josephson junctions.\cite{Cuevas07,Bergeret08} In the
junctions with the lower doped nanowires the Josephson supercurrent
$I_c$ as well as the conductance $G_S$ fluctuates when the gate voltage
is varied. The measured average amplitude of supercurrent fluctuations
was smaller than the theoretically expected value. The large difference
between both values is attributed to the presence of a barrier at the
Nb/nanowire interface. The average conductance fluctuation amplitude
for the Nb/InAs-nanowire samples was considerably larger than the
corresponding value for a reference sample with normal conducting Au/Ti
contacts. We attribute this enhancement to the contribution of
phase-coherent Andreev reflection.

\subsection{Acknowledgements}

The authors are grateful to H. Kertz for assistance during the
measurements and S. Trellenkamp for electron beam writing. H.Y.G.
thanks The Scientific and Technological Research Council of Turkey
(TUBITAK) foundation. I.E.B. acknowledges the Russian Foundation for
Basic Research, Project No. RFBR 09-02-01499  for financial support.


\begin{thebibliography}{46}
\expandafter\ifx\csname natexlab\endcsname\relax\def\natexlab#1{#1}\fi
\expandafter\ifx\csname bibnamefont\endcsname\relax
  \def\bibnamefont#1{#1}\fi
\expandafter\ifx\csname bibfnamefont\endcsname\relax
  \def\bibfnamefont#1{#1}\fi
\expandafter\ifx\csname citenamefont\endcsname\relax
  \def\citenamefont#1{#1}\fi
\expandafter\ifx\csname url\endcsname\relax
  \def\url#1{\texttt{#1}}\fi
\expandafter\ifx\csname urlprefix\endcsname\relax\def\urlprefix{URL
}\fi \providecommand{\bibinfo}[2]{#2}
\providecommand{\eprint}[2][]{\url{#2}}

\bibitem[{\citenamefont{De~Franceschi
    et~al.}(2010)\citenamefont{De~Franceschi,
  Kouwenhoven, Sch\"onenberger, and Wernsdorfer}}]{DeFranceschi10}
\bibinfo{author}{\bibfnamefont{S.}~\bibnamefont{De~Franceschi}},
  \bibinfo{author}{\bibfnamefont{L.}~\bibnamefont{Kouwenhoven}},
  \bibinfo{author}{\bibfnamefont{C.}~\bibnamefont{Sch\"onenberger}},
  \bibnamefont{and}
  \bibinfo{author}{\bibfnamefont{W.}~\bibnamefont{Wernsdorfer}},
  \bibinfo{journal}{Nature Nano} \textbf{\bibinfo{volume}{5}},
  \bibinfo{pages}{703} (\bibinfo{year}{2010}).

\bibitem[{\citenamefont{Sch\"{a}pers}(2001)}]{Schaepers01}
    \bibinfo{author}{\bibfnamefont{T.}~\bibnamefont{Sch\"{a}pers}},
  \emph{\bibinfo{title}{Superconductor/Semiconductor Junctions}}, vol.
  \bibinfo{volume}{174} (\bibinfo{publisher}{Springer Tracts on Modern
  Physics}, \bibinfo{year}{2001}).

\bibitem[{\citenamefont{Akazaki et~al.}(1996)\citenamefont{Akazaki,
    Takayanagi,
  Nitta, and Enoki}}]{Akazaki96}
\bibinfo{author}{\bibfnamefont{T.}~\bibnamefont{Akazaki}},
  \bibinfo{author}{\bibfnamefont{H.}~\bibnamefont{Takayanagi}},
  \bibinfo{author}{\bibfnamefont{J.}~\bibnamefont{Nitta}}, \bibnamefont{and}
  \bibinfo{author}{\bibfnamefont{T.}~\bibnamefont{Enoki}},
  \bibinfo{journal}{Appl. Phys. Lett.} \textbf{\bibinfo{volume}{68}},
  \bibinfo{pages}{418} (\bibinfo{year}{1996}).

\bibitem[{\citenamefont{Takayanagi
  et~al.}(1995{\natexlab{a}})\citenamefont{Takayanagi, Akazaki, and
  Nitta}}]{Hideaki95}
\bibinfo{author}{\bibfnamefont{H.}~\bibnamefont{Takayanagi}},
  \bibinfo{author}{\bibfnamefont{T.}~\bibnamefont{Akazaki}}, \bibnamefont{and}
  \bibinfo{author}{\bibfnamefont{J.}~\bibnamefont{Nitta}},
  \bibinfo{journal}{Phys. Rev. Lett.} \textbf{\bibinfo{volume}{75}},
  \bibinfo{pages}{3533} (\bibinfo{year}{1995}{\natexlab{a}}).

\bibitem[{\citenamefont{Sch\"apers
    et~al.}(2003)\citenamefont{Sch\"apers,
  Guzenko, M\"uller, Golubov, Brinkman, Crecelius, Kaluza, and
  L\"uth}}]{Schaepers03a}
\bibinfo{author}{\bibfnamefont{Th.}~\bibnamefont{Sch\"apers}},
  \bibinfo{author}{\bibfnamefont{V.~A.} \bibnamefont{Guzenko}},
  \bibinfo{author}{\bibfnamefont{R.~P.} \bibnamefont{M\"uller}},
  \bibinfo{author}{\bibfnamefont{A.~A.} \bibnamefont{Golubov}},
  \bibinfo{author}{\bibfnamefont{A.}~\bibnamefont{Brinkman}},
  \bibinfo{author}{\bibfnamefont{G.}~\bibnamefont{Crecelius}},
  \bibinfo{author}{\bibfnamefont{A.}~\bibnamefont{Kaluza}}, \bibnamefont{and}
  \bibinfo{author}{\bibfnamefont{H.}~\bibnamefont{L\"uth}},
  \bibinfo{journal}{Phys. Rev. B} \textbf{\bibinfo{volume}{67}},
  \bibinfo{pages}{014522} (\bibinfo{year}{2003}).

\bibitem[{\citenamefont{Doh et~al.}(2005)\citenamefont{Doh, van Dam,
    Roest,
  Bakkers, Kouwenhoven, and Franceschi}}]{Doh05}
\bibinfo{author}{\bibfnamefont{Y.-J.} \bibnamefont{Doh}},
  \bibinfo{author}{\bibfnamefont{J.~A.} \bibnamefont{van Dam}},
  \bibinfo{author}{\bibfnamefont{A.~L.} \bibnamefont{Roest}},
  \bibinfo{author}{\bibfnamefont{E.~P. A.~M.} \bibnamefont{Bakkers}},
  \bibinfo{author}{\bibfnamefont{L.~P.} \bibnamefont{Kouwenhoven}},
  \bibnamefont{and} \bibinfo{author}{\bibfnamefont{S.~D.}
  \bibnamefont{Franceschi}}, \bibinfo{journal}{Science}
  \textbf{\bibinfo{volume}{309}}, \bibinfo{pages}{272} (\bibinfo{year}{2005}).

\bibitem[{\citenamefont{Xiang et~al.}(2006)\citenamefont{Xiang, Vidan,
    Tinkham,
  Westervelt, and Lieber}}]{Xiang06}
\bibinfo{author}{\bibfnamefont{J.}~\bibnamefont{Xiang}},
  \bibinfo{author}{\bibfnamefont{A.}~\bibnamefont{Vidan}},
  \bibinfo{author}{\bibfnamefont{M.}~\bibnamefont{Tinkham}},
  \bibinfo{author}{\bibfnamefont{R.~M.} \bibnamefont{Westervelt}},
  \bibnamefont{and} \bibinfo{author}{\bibfnamefont{C.~M.}
  \bibnamefont{Lieber}}, \bibinfo{journal}{Nature Nanotechnology}
  \textbf{\bibinfo{volume}{1}}, \bibinfo{pages}{208} (\bibinfo{year}{2006}).

\bibitem[{\citenamefont{van Dam et~al.}(2006)\citenamefont{van Dam,
    Nazarov,
  Bakkers, De~Franceschi, and Kouwenhoven}}]{vanDam06}
\bibinfo{author}{\bibfnamefont{J.~A.} \bibnamefont{van Dam}},
  \bibinfo{author}{\bibfnamefont{Y.~V.} \bibnamefont{Nazarov}},
  \bibinfo{author}{\bibfnamefont{E.~P. A.~M.} \bibnamefont{Bakkers}},
  \bibinfo{author}{\bibfnamefont{S.}~\bibnamefont{De~Franceschi}},
  \bibnamefont{and} \bibinfo{author}{\bibfnamefont{L.~P.}
  \bibnamefont{Kouwenhoven}}, \bibinfo{journal}{Nature}
  \textbf{\bibinfo{volume}{442}}, \bibinfo{pages}{667} (\bibinfo{year}{2006}).

\bibitem[{\citenamefont{Frielinghaus
    et~al.}(2010)\citenamefont{Frielinghaus,
  Batov, Weides, Kohlstedt, Calarco, and Sch\"{a}pers}}]{Frielinghaus10}
\bibinfo{author}{\bibfnamefont{R.}~\bibnamefont{Frielinghaus}},
  \bibinfo{author}{\bibfnamefont{I.~E.} \bibnamefont{Batov}},
  \bibinfo{author}{\bibfnamefont{M.}~\bibnamefont{Weides}},
  \bibinfo{author}{\bibfnamefont{H.}~\bibnamefont{Kohlstedt}},
  \bibinfo{author}{\bibfnamefont{R.}~\bibnamefont{Calarco}}, \bibnamefont{and}
  \bibinfo{author}{\bibfnamefont{Th.}~\bibnamefont{Sch\"{a}pers}},
  \bibinfo{journal}{Appl. Phys. Lett.} \textbf{\bibinfo{volume}{96}},
  \bibinfo{pages}{132504} (\bibinfo{year}{2010}).

\bibitem[{\citenamefont{Roddaro et~al.}(2011)\citenamefont{Roddaro,
    Pescaglini,
  Ercolani, Sorba, Giazotto, and Beltram}}]{Roddaro11}
\bibinfo{author}{\bibfnamefont{S.}~\bibnamefont{Roddaro}},
  \bibinfo{author}{\bibfnamefont{A.}~\bibnamefont{Pescaglini}},
  \bibinfo{author}{\bibfnamefont{D.}~\bibnamefont{Ercolani}},
  \bibinfo{author}{\bibfnamefont{L.}~\bibnamefont{Sorba}},
  \bibinfo{author}{\bibfnamefont{F.}~\bibnamefont{Giazotto}}, \bibnamefont{and}
  \bibinfo{author}{\bibfnamefont{F.}~\bibnamefont{Beltram}},
  \bibinfo{journal}{Nano Research} \textbf{\bibinfo{volume}{4}},
  \bibinfo{pages}{259} (\bibinfo{year}{2011}).

\bibitem[{\citenamefont{Nishio et~al.}(2011)\citenamefont{Nishio,
    Kozakai,
  Amaha, Larsson, Nilsson, Xu, Zhang, Tateno, Takayanagi, and
  Ishibashi}}]{Nishio11}
\bibinfo{author}{\bibfnamefont{T.}~\bibnamefont{Nishio}},
  \bibinfo{author}{\bibfnamefont{T.}~\bibnamefont{Kozakai}},
  \bibinfo{author}{\bibfnamefont{S.}~\bibnamefont{Amaha}},
  \bibinfo{author}{\bibfnamefont{M.}~\bibnamefont{Larsson}},
  \bibinfo{author}{\bibfnamefont{H.~A.} \bibnamefont{Nilsson}},
  \bibinfo{author}{\bibfnamefont{H.~Q.} \bibnamefont{Xu}},
  \bibinfo{author}{\bibfnamefont{G.}~\bibnamefont{Zhang}},
  \bibinfo{author}{\bibfnamefont{K.}~\bibnamefont{Tateno}},
  \bibinfo{author}{\bibfnamefont{H.}~\bibnamefont{Takayanagi}},
  \bibnamefont{and}
  \bibinfo{author}{\bibfnamefont{K.}~\bibnamefont{Ishibashi}},
  \bibinfo{journal}{Nanotechnology} \textbf{\bibinfo{volume}{22}},
  \bibinfo{pages}{445701} (\bibinfo{year}{2011}).

\bibitem[{\citenamefont{Smit et~al.}(1989)\citenamefont{Smit, Koenders,
    and
  M\"onch}}]{Smit89}
\bibinfo{author}{\bibfnamefont{K.}~\bibnamefont{Smit}},
  \bibinfo{author}{\bibfnamefont{L.}~\bibnamefont{Koenders}}, \bibnamefont{and}
  \bibinfo{author}{\bibfnamefont{W.}~\bibnamefont{M\"onch}},
  \bibinfo{journal}{Journal of Vacuum Science \& Technology B (Microelectronics
  Processing and Phenomena)} \textbf{\bibinfo{volume}{7}}, \bibinfo{pages}{888
  } (\bibinfo{year}{1989}).

\bibitem[{\citenamefont{Sand-Jespersen
  et~al.}(2007)\citenamefont{Sand-Jespersen, Paaske, Andersen, Grove-Rasmussen,
  J\o{}rgensen, Aagesen, S\o{}rensen, Lindelof, Flensberg, and
  Nyg\aa{}rd}}]{Sand-Jespersen07}
\bibinfo{author}{\bibfnamefont{T.}~\bibnamefont{Sand-Jespersen}},
  \bibinfo{author}{\bibfnamefont{J.}~\bibnamefont{Paaske}},
  \bibinfo{author}{\bibfnamefont{B.~M.} \bibnamefont{Andersen}},
  \bibinfo{author}{\bibfnamefont{K.}~\bibnamefont{Grove-Rasmussen}},
  \bibinfo{author}{\bibfnamefont{H.~I.} \bibnamefont{J\o{}rgensen}},
  \bibinfo{author}{\bibfnamefont{M.}~\bibnamefont{Aagesen}},
  \bibinfo{author}{\bibfnamefont{C.~B.} \bibnamefont{S\o{}rensen}},
  \bibinfo{author}{\bibfnamefont{P.~E.} \bibnamefont{Lindelof}},
  \bibinfo{author}{\bibfnamefont{K.}~\bibnamefont{Flensberg}},
  \bibnamefont{and}
  \bibinfo{author}{\bibfnamefont{J.}~\bibnamefont{Nyg\aa{}rd}},
  \bibinfo{journal}{Phys. Rev. Lett.} \textbf{\bibinfo{volume}{99}},
  \bibinfo{pages}{126603} (\bibinfo{year}{2007}).

\bibitem[{\citenamefont{Hofstetter
    et~al.}(2009)\citenamefont{Hofstetter,
  Csonka, Nygard, and Sch\"onenberger}}]{Hofstetter09}
\bibinfo{author}{\bibfnamefont{L.}~\bibnamefont{Hofstetter}},
  \bibinfo{author}{\bibfnamefont{S.}~\bibnamefont{Csonka}},
  \bibinfo{author}{\bibfnamefont{J.}~\bibnamefont{Nygard}}, \bibnamefont{and}
  \bibinfo{author}{\bibfnamefont{C.}~\bibnamefont{Sch\"onenberger}},
  \bibinfo{journal}{Nature} \textbf{\bibinfo{volume}{461}},
  \bibinfo{pages}{960} (\bibinfo{year}{2009}).

\bibitem[{\citenamefont{Spathis et~al.}(2011)\citenamefont{Spathis,
    Biswas,
  Roddaro, Sorba, Giazotto, and Beltram}}]{Spathis11}
\bibinfo{author}{\bibfnamefont{P.}~\bibnamefont{Spathis}},
  \bibinfo{author}{\bibfnamefont{S.}~\bibnamefont{Biswas}},
  \bibinfo{author}{\bibfnamefont{S.}~\bibnamefont{Roddaro}},
  \bibinfo{author}{\bibfnamefont{L.}~\bibnamefont{Sorba}},
  \bibinfo{author}{\bibfnamefont{F.}~\bibnamefont{Giazotto}}, \bibnamefont{and}
  \bibinfo{author}{\bibfnamefont{F.}~\bibnamefont{Beltram}},
  \bibinfo{journal}{Nanotechnology} \textbf{\bibinfo{volume}{22}},
  \bibinfo{pages}{105201} (\bibinfo{year}{2011}).

\bibitem[{\citenamefont{Fasth et~al.}(2007)\citenamefont{Fasth, Fuhrer,
  Samuelson, Golovach, and Loss}}]{Fasth07}
\bibinfo{author}{\bibfnamefont{C.}~\bibnamefont{Fasth}},
  \bibinfo{author}{\bibfnamefont{A.}~\bibnamefont{Fuhrer}},
  \bibinfo{author}{\bibfnamefont{L.}~\bibnamefont{Samuelson}},
  \bibinfo{author}{\bibfnamefont{V.~N.} \bibnamefont{Golovach}},
  \bibnamefont{and} \bibinfo{author}{\bibfnamefont{D.}~\bibnamefont{Loss}},
  \bibinfo{journal}{Phys. Rev. Lett.} \textbf{\bibinfo{volume}{98}},
  \bibinfo{pages}{266801} (\bibinfo{year}{2007}).

\bibitem[{\citenamefont{Lutchyn et~al.}(2010)\citenamefont{Lutchyn,
    Sau, and
  Das~Sarma}}]{Lutchyn10}
\bibinfo{author}{\bibfnamefont{R.~M.} \bibnamefont{Lutchyn}},
  \bibinfo{author}{\bibfnamefont{J.~D.} \bibnamefont{Sau}}, \bibnamefont{and}
  \bibinfo{author}{\bibfnamefont{S.}~\bibnamefont{Das~Sarma}},
  \bibinfo{journal}{Phys. Rev. Lett.} \textbf{\bibinfo{volume}{105}},
  \bibinfo{pages}{077001} (\bibinfo{year}{2010}).

\bibitem[{\citenamefont{Cuevas and Bergeret}(2007)}]{Cuevas07}
    \bibinfo{author}{\bibfnamefont{J.~C.} \bibnamefont{Cuevas}}
    \bibnamefont{and}
  \bibinfo{author}{\bibfnamefont{F.~S.} \bibnamefont{Bergeret}},
  \bibinfo{journal}{Physical Review Letters} \textbf{\bibinfo{volume}{99}},
  \bibinfo{eid}{217002} (pages~\bibinfo{numpages}{4}) (\bibinfo{year}{2007}).

\bibitem[{\citenamefont{Bergeret and Cuevas}(2008)}]{Bergeret08}
    \bibinfo{author}{\bibfnamefont{F.~S.} \bibnamefont{Bergeret}}
    \bibnamefont{and}
  \bibinfo{author}{\bibfnamefont{J.~C.} \bibnamefont{Cuevas}},
  \bibinfo{journal}{J. Low Temp. Phys.} \textbf{\bibinfo{volume}{153}},
  \bibinfo{pages}{304} (\bibinfo{year}{2008}).

\bibitem[{\citenamefont{Angers et~al.}(2008)\citenamefont{Angers,
    Chiodi,
  Montambaux, Ferrier, Gu\'{e}ron, Bouchiat, and Cuevas}}]{Angers08}
\bibinfo{author}{\bibfnamefont{L.}~\bibnamefont{Angers}},
  \bibinfo{author}{\bibfnamefont{F.}~\bibnamefont{Chiodi}},
  \bibinfo{author}{\bibfnamefont{G.}~\bibnamefont{Montambaux}},
  \bibinfo{author}{\bibfnamefont{M.}~\bibnamefont{Ferrier}},
  \bibinfo{author}{\bibfnamefont{S.}~\bibnamefont{Gu\'{e}ron}},
  \bibinfo{author}{\bibfnamefont{H.}~\bibnamefont{Bouchiat}}, \bibnamefont{and}
  \bibinfo{author}{\bibfnamefont{J.~C.} \bibnamefont{Cuevas}},
  \bibinfo{journal}{Physical Review B (Condensed Matter and Materials Physics)}
  \textbf{\bibinfo{volume}{77}}, \bibinfo{eid}{165408}
  (pages~\bibinfo{numpages}{12}) (\bibinfo{year}{2008}).

\bibitem[{\citenamefont{Doh et~al.}(2008)\citenamefont{Doh, Franceschi,
  Bakkers, and Kouwenhoven}}]{Doh08}
\bibinfo{author}{\bibfnamefont{Y.-J.} \bibnamefont{Doh}},
  \bibinfo{author}{\bibfnamefont{S.~D.} \bibnamefont{Franceschi}},
  \bibinfo{author}{\bibfnamefont{E.~P. A.~M.} \bibnamefont{Bakkers}},
  \bibnamefont{and} \bibinfo{author}{\bibfnamefont{L.~P.}
  \bibnamefont{Kouwenhoven}}, \bibinfo{journal}{Nano Letters}
  \textbf{\bibinfo{volume}{8}}, \bibinfo{pages}{4098} (\bibinfo{year}{2008}).

\bibitem[{\citenamefont{Jespersen et~al.}(2009)\citenamefont{Jespersen,
  Polianski, Sørensen, Flensberg, and Nygård}}]{Jespersen09}
\bibinfo{author}{\bibfnamefont{T.~S.} \bibnamefont{Jespersen}},
  \bibinfo{author}{\bibfnamefont{M.~L.} \bibnamefont{Polianski}},
  \bibinfo{author}{\bibfnamefont{C.~B.} \bibnamefont{Sørensen}},
  \bibinfo{author}{\bibfnamefont{K.}~\bibnamefont{Flensberg}},
  \bibnamefont{and} \bibinfo{author}{\bibfnamefont{J.}~\bibnamefont{Nygård}},
  \bibinfo{journal}{New Journal of Physics} \textbf{\bibinfo{volume}{11}},
  \bibinfo{pages}{113025} (\bibinfo{year}{2009}).

\bibitem[{\citenamefont{Takayanagi
  et~al.}(1995{\natexlab{b}})\citenamefont{Takayanagi, Hansen, and
  Nitta}}]{Takayanagi95}
\bibinfo{author}{\bibfnamefont{H.}~\bibnamefont{Takayanagi}},
  \bibinfo{author}{\bibfnamefont{J.~B.} \bibnamefont{Hansen}},
  \bibnamefont{and} \bibinfo{author}{\bibfnamefont{J.}~\bibnamefont{Nitta}},
  \bibinfo{journal}{Phys. Rev. Lett.} \textbf{\bibinfo{volume}{74}},
  \bibinfo{pages}{166} (\bibinfo{year}{1995}{\natexlab{b}}).

\bibitem[{\citenamefont{Wirths et~al.}(2011)\citenamefont{Wirths, Weis,
    Winden,
  Sladek, Volk, Alagha, Weirich, von~der Ahe, Hardtdegen, L\"{u}th
  et~al.}}]{Wirths11}
\bibinfo{author}{\bibfnamefont{S.}~\bibnamefont{Wirths}},
  \bibinfo{author}{\bibfnamefont{K.}~\bibnamefont{Weis}},
  \bibinfo{author}{\bibfnamefont{A.}~\bibnamefont{Winden}},
  \bibinfo{author}{\bibfnamefont{K.}~\bibnamefont{Sladek}},
  \bibinfo{author}{\bibfnamefont{C.}~\bibnamefont{Volk}},
  \bibinfo{author}{\bibfnamefont{S.}~\bibnamefont{Alagha}},
  \bibinfo{author}{\bibfnamefont{T.~E.} \bibnamefont{Weirich}},
  \bibinfo{author}{\bibfnamefont{M.}~\bibnamefont{von~der Ahe}},
  \bibinfo{author}{\bibfnamefont{H.}~\bibnamefont{Hardtdegen}},
  \bibinfo{author}{\bibfnamefont{H.}~\bibnamefont{L\"{u}th}},
  \bibnamefont{et~al.}, \bibinfo{journal}{Journal of Applied Physics}
  \textbf{\bibinfo{volume}{110}}, \bibinfo{eid}{053709} (\bibinfo{year}{2011}).

\bibitem[{\citenamefont{Jarillo-Herrero
  et~al.}(2006)\citenamefont{Jarillo-Herrero, van Dam, and
  Kouwenhoven}}]{Jarillo-Herrero06}
\bibinfo{author}{\bibfnamefont{P.}~\bibnamefont{Jarillo-Herrero}},
  \bibinfo{author}{\bibfnamefont{J.}~\bibnamefont{van Dam}}, \bibnamefont{and}
  \bibinfo{author}{\bibfnamefont{L.}~\bibnamefont{Kouwenhoven}},
  \bibinfo{journal}{Nature} \textbf{\bibinfo{volume}{439}},
  \bibinfo{pages}{953} (\bibinfo{year}{2006}).

\bibitem[{\citenamefont{Pallecchi et~al.}(2008)\citenamefont{Pallecchi,
    Gaa\ss,
  Ryndyk, and Strunk}}]{Pallecchi08}
\bibinfo{author}{\bibfnamefont{E.}~\bibnamefont{Pallecchi}},
  \bibinfo{author}{\bibfnamefont{M.}~\bibnamefont{Gaa\ss}},
  \bibinfo{author}{\bibfnamefont{D.~A.} \bibnamefont{Ryndyk}},
  \bibnamefont{and} \bibinfo{author}{\bibfnamefont{C.}~\bibnamefont{Strunk}},
  \bibinfo{journal}{Applied Physics Letters} \textbf{\bibinfo{volume}{93}},
  \bibinfo{eid}{072501} (\bibinfo{year}{2008}).

\bibitem[{\citenamefont{Courtois et~al.}(2008)\citenamefont{Courtois,
    Meschke,
  Peltonen, and Pekola}}]{Courtois08}
\bibinfo{author}{\bibfnamefont{H.}~\bibnamefont{Courtois}},
  \bibinfo{author}{\bibfnamefont{M.}~\bibnamefont{Meschke}},
  \bibinfo{author}{\bibfnamefont{J.~T.} \bibnamefont{Peltonen}},
  \bibnamefont{and} \bibinfo{author}{\bibfnamefont{J.~P.}
  \bibnamefont{Pekola}}, \bibinfo{journal}{Phys. Rev. Lett.}
  \textbf{\bibinfo{volume}{101}}, \bibinfo{eid}{067002}
 (\bibinfo{year}{2008}).

\bibitem[{\citenamefont{Octavio et~al.}(1983)\citenamefont{Octavio,
    Tinkham,
  Blonder, and Klapwijk}}]{Octavio1983}
\bibinfo{author}{\bibfnamefont{M.}~\bibnamefont{Octavio}},
  \bibinfo{author}{\bibfnamefont{M.}~\bibnamefont{Tinkham}},
  \bibinfo{author}{\bibfnamefont{G.~E.} \bibnamefont{Blonder}},
  \bibnamefont{and} \bibinfo{author}{\bibfnamefont{T.~M.}
  \bibnamefont{Klapwijk}}, \bibinfo{journal}{Phys. Rev. B}
  \textbf{\bibinfo{volume}{27}}, \bibinfo{pages}{6739} (\bibinfo{year}{1983}).

\bibitem[{\citenamefont{Flensberg et~al.}(1988)\citenamefont{Flensberg,
    Hansen,
  and Octavio}}]{Flensberg88}
\bibinfo{author}{\bibfnamefont{K.}~\bibnamefont{Flensberg}},
  \bibinfo{author}{\bibfnamefont{J.~B.} \bibnamefont{Hansen}},
  \bibnamefont{and} \bibinfo{author}{\bibfnamefont{M.}~\bibnamefont{Octavio}},
  \bibinfo{journal}{Phys. Rev. B} \textbf{\bibinfo{volume}{38}},
  \bibinfo{pages}{8707} (\bibinfo{year}{1988}).

\bibitem[{\citenamefont{Cuevas et~al.}(2006)\citenamefont{Cuevas,
    Hammer, Kopu,
  Viljas, and Eschrig}}]{Cuevas06}
\bibinfo{author}{\bibfnamefont{J.~C.} \bibnamefont{Cuevas}},
  \bibinfo{author}{\bibfnamefont{J.}~\bibnamefont{Hammer}},
  \bibinfo{author}{\bibfnamefont{J.}~\bibnamefont{Kopu}},
  \bibinfo{author}{\bibfnamefont{J.~K.} \bibnamefont{Viljas}},
  \bibnamefont{and} \bibinfo{author}{\bibfnamefont{M.}~\bibnamefont{Eschrig}},
  \bibinfo{journal}{Phys. Rev. B} \textbf{\bibinfo{volume}{73}},
  \bibinfo{pages}{184505} (\bibinfo{year}{2006}).

\bibitem[{\citenamefont{Blonder et~al.}(1982)\citenamefont{Blonder,
    Tinkham,
  and Klapwijk}}]{Blonder82}
\bibinfo{author}{\bibfnamefont{G.~E.} \bibnamefont{Blonder}},
  \bibinfo{author}{\bibfnamefont{M.}~\bibnamefont{Tinkham}}, \bibnamefont{and}
  \bibinfo{author}{\bibfnamefont{T.~M.} \bibnamefont{Klapwijk}},
  \bibinfo{journal}{Phys. Rev. B} \textbf{\bibinfo{volume}{25}},
  \bibinfo{pages}{4515} (\bibinfo{year}{1982}).

\bibitem[{\citenamefont{Batov et~al.}(2004)\citenamefont{Batov,
    Sch\"{a}pers,
  Golubov, and Ustinov}}]{Batov04}
\bibinfo{author}{\bibfnamefont{I.~E.} \bibnamefont{Batov}},
  \bibinfo{author}{\bibfnamefont{Th.}~\bibnamefont{Sch\"{a}pers}},
  \bibinfo{author}{\bibfnamefont{A.~A.} \bibnamefont{Golubov}},
  \bibnamefont{and} \bibinfo{author}{\bibfnamefont{A.~V.}
  \bibnamefont{Ustinov}}, \bibinfo{journal}{Journal of Applied Physics}
  \textbf{\bibinfo{volume}{96}}, \bibinfo{pages}{3366} (\bibinfo{year}{2004}).

\bibitem[{\citenamefont{Chrestin et~al.}(1997)\citenamefont{Chrestin,
  Matsuyama, and Merkt}}]{Chrestin97}
\bibinfo{author}{\bibfnamefont{A.}~\bibnamefont{Chrestin}},
  \bibinfo{author}{\bibfnamefont{T.}~\bibnamefont{Matsuyama}},
  \bibnamefont{and} \bibinfo{author}{\bibfnamefont{U.}~\bibnamefont{Merkt}},
  \bibinfo{journal}{Phys. Rev. B} \textbf{\bibinfo{volume}{55}},
  \bibinfo{pages}{8457} (\bibinfo{year}{1997}).

\bibitem[{\citenamefont{Carbotte}(1990)}]{Carbotte90}
    \bibinfo{author}{\bibfnamefont{J.~P.} \bibnamefont{Carbotte}},
  \bibinfo{journal}{Rev. Mod. Phys.} \textbf{\bibinfo{volume}{62}},
  \bibinfo{pages}{1027} (\bibinfo{year}{1990}).

\bibitem[{\citenamefont{Hammer et~al.}(2007)\citenamefont{Hammer,
    Cuevas,
  Bergeret, and Belzig}}]{Hammer07}
\bibinfo{author}{\bibfnamefont{J.~C.} \bibnamefont{Hammer}},
  \bibinfo{author}{\bibfnamefont{J.~C.} \bibnamefont{Cuevas}},
  \bibinfo{author}{\bibfnamefont{F.~S.} \bibnamefont{Bergeret}},
  \bibnamefont{and} \bibinfo{author}{\bibfnamefont{W.}~\bibnamefont{Belzig}},
  \bibinfo{journal}{Physical Review B} \textbf{\bibinfo{volume}{76}},
  \bibinfo{eid}{064514} (\bibinfo{year}{2007}).

\bibitem[{\citenamefont{Mac\^edo and Chalker}(1994)}]{Macedo94}
    \bibinfo{author}{\bibfnamefont{A.~M.~S.} \bibnamefont{Mac\^edo}}
  \bibnamefont{and} \bibinfo{author}{\bibfnamefont{J.~T.}
  \bibnamefont{Chalker}}, \bibinfo{journal}{Phys. Rev. B}
  \textbf{\bibinfo{volume}{49}}, \bibinfo{pages}{4695} (\bibinfo{year}{1994}).

\bibitem[{\citenamefont{Beenakker}(1991)}]{Beenakker91a}
    \bibinfo{author}{\bibfnamefont{C.~W.~J.} \bibnamefont{Beenakker}},
  \bibinfo{journal}{Phys. Rev. Lett} \textbf{\bibinfo{volume}{67}},
  \bibinfo{pages}{3836} (\bibinfo{year}{1991}).

\bibitem[{\citenamefont{Beenakker and Rajaei}(1994)}]{Beenakker94}
    \bibinfo{author}{\bibfnamefont{C.~W.~J.} \bibnamefont{Beenakker}}
  \bibnamefont{and} \bibinfo{author}{\bibfnamefont{B.}~\bibnamefont{Rajaei}},
  \bibinfo{journal}{Phys. Rev. B} \textbf{\bibinfo{volume}{49}},
  \bibinfo{pages}{7499} (\bibinfo{year}{1994}).

\bibitem[{\citenamefont{Al'tshuler and Spivak}(1987)}]{Altshuler87}
    \bibinfo{author}{\bibfnamefont{B.}~\bibnamefont{Al'tshuler}}
    \bibnamefont{and}
  \bibinfo{author}{\bibfnamefont{B.}~\bibnamefont{Spivak}},
  \bibinfo{journal}{Zh. Eksp. Teo. Fiz. [Sov. Physics-JETP {\bf 65}, 343-347
  (1987)]} \textbf{\bibinfo{volume}{92}}, \bibinfo{pages}{609}
  (\bibinfo{year}{1987}).

\bibitem[{\citenamefont{Houzet and Skvortsov}(2008)}]{Houyet08}
    \bibinfo{author}{\bibfnamefont{M.}~\bibnamefont{Houzet}}
    \bibnamefont{and}
  \bibinfo{author}{\bibfnamefont{M.~A.} \bibnamefont{Skvortsov}},
  \bibinfo{journal}{Phys. Rev. B} \textbf{\bibinfo{volume}{77}},
  \bibinfo{pages}{024525} (\bibinfo{year}{2008}).

\bibitem[{\citenamefont{Sch\"apers
    et~al.}(1997)\citenamefont{Sch\"apers,
  Kaluza, Neurohr, Malindretos, Crecelius, van~der Hart, Hardtdegen, and
  L\"uth}}]{Schaepers97}
\bibinfo{author}{\bibfnamefont{Th.}~\bibnamefont{Sch\"apers}},
  \bibinfo{author}{\bibfnamefont{A.}~\bibnamefont{Kaluza}},
  \bibinfo{author}{\bibfnamefont{K.}~\bibnamefont{Neurohr}},
  \bibinfo{author}{\bibfnamefont{J.}~\bibnamefont{Malindretos}},
  \bibinfo{author}{\bibfnamefont{G.}~\bibnamefont{Crecelius}},
  \bibinfo{author}{\bibfnamefont{A.}~\bibnamefont{van~der Hart}},
  \bibinfo{author}{\bibfnamefont{H.}~\bibnamefont{Hardtdegen}},
  \bibnamefont{and} \bibinfo{author}{\bibfnamefont{H.}~\bibnamefont{L\"uth}},
  \bibinfo{journal}{Appl. Phys. Lett.} \textbf{\bibinfo{volume}{71}},
  \bibinfo{pages}{3575} (\bibinfo{year}{1997}).

\bibitem[{\citenamefont{Stone}(1985)}]{Stone85}
    \bibinfo{author}{\bibfnamefont{A.~D.} \bibnamefont{Stone}},
  \bibinfo{journal}{Phys. Rev. Lett.} \textbf{\bibinfo{volume}{54}},
  \bibinfo{pages}{2692} (\bibinfo{year}{1985}).

\bibitem[{\citenamefont{Al'tshuler}(1985)}]{Altshuler85b}
    \bibinfo{author}{\bibfnamefont{B.}~\bibnamefont{Al'tshuler}},
  \bibinfo{journal}{Pis'ma Zh. Eksp. Teo. Fiz. [JETP Lett. {\bf 41}, 648-651
  (1985)]} \textbf{\bibinfo{volume}{41}}, \bibinfo{pages}{530}
  (\bibinfo{year}{1985}).

\bibitem[{\citenamefont{Bl\"omers et~al.}(2011)\citenamefont{Bl\"omers,
    Lepsa,
  Luysberg, Gr\"utzmacher, L\"uth, and Sch\"apers}}]{Bloemers11}
\bibinfo{author}{\bibfnamefont{C.}~\bibnamefont{Bl\"omers}},
  \bibinfo{author}{\bibfnamefont{M.~I.} \bibnamefont{Lepsa}},
  \bibinfo{author}{\bibfnamefont{M.}~\bibnamefont{Luysberg}},
  \bibinfo{author}{\bibfnamefont{D.}~\bibnamefont{Gr\"utzmacher}},
  \bibinfo{author}{\bibfnamefont{H.}~\bibnamefont{L\"uth}}, \bibnamefont{and}
  \bibinfo{author}{\bibfnamefont{Th.}~\bibnamefont{Sch\"apers}},
  \bibinfo{journal}{Nano Letters} \textbf{\bibinfo{volume}{11}},
  \bibinfo{pages}{3550} (\bibinfo{year}{2011}).

\bibitem[{\citenamefont{Trbovic et~al.}(2010)\citenamefont{Trbovic,
    Minder,
  Freitag, and Sch\"onenberger}}]{Trbovic10}
\bibinfo{author}{\bibfnamefont{J.}~\bibnamefont{Trbovic}},
  \bibinfo{author}{\bibfnamefont{N.}~\bibnamefont{Minder}},
  \bibinfo{author}{\bibfnamefont{F.}~\bibnamefont{Freitag}}, \bibnamefont{and}
  \bibinfo{author}{\bibfnamefont{C.}~\bibnamefont{Sch\"onenberger}},
  \bibinfo{journal}{Nanotechnology} \textbf{\bibinfo{volume}{21}},
  \bibinfo{pages}{274005} (\bibinfo{year}{2010}).

\bibitem[{\citenamefont{Ojeda-Aristizabal
  et~al.}(2009)\citenamefont{Ojeda-Aristizabal, Ferrier, Gu\'eron, and
  Bouchiat}}]{Ojeda-Aristizabal09}
\bibinfo{author}{\bibfnamefont{C.}~\bibnamefont{Ojeda-Aristizabal}},
  \bibinfo{author}{\bibfnamefont{M.}~\bibnamefont{Ferrier}},
  \bibinfo{author}{\bibfnamefont{S.}~\bibnamefont{Gu\'eron}}, \bibnamefont{and}
  \bibinfo{author}{\bibfnamefont{H.}~\bibnamefont{Bouchiat}},
  \bibinfo{journal}{Phys. Rev. B} \textbf{\bibinfo{volume}{79}},
  \bibinfo{pages}{165436} (\bibinfo{year}{2009}).

\end{thebibliography}

\end{document}